\documentclass[aps,prb,showpacs,floatfix,twocolumn,10pt]{revtex4-1}

\usepackage{graphicx}
\usepackage{amsmath}
\usepackage{amssymb}
\usepackage[caption=false]{subfig}

\newcommand{\bne}{\begin{equation*}}
\newcommand{\ede}{\end{equation*}}
\newcommand{\bea}{\begin{eqnarray*}}
\newcommand{\eea}{\end{eqnarray*}}
\newcommand{\bnen}{\begin{equation}}
\newcommand{\eden}{\end{equation}}
\newcommand{\bnsn}{\begin{subequations}}
\newcommand{\edsn}{\end{subequations}}
\newcommand{\bean}{\begin{eqnarray}}
\newcommand{\eean}{\end{eqnarray}}
\newcommand{\bna}{\begin{array}}
\newcommand{\eda}{\end{array}}

\renewcommand{\vec}[1]{\mbox{\boldmath{$#1$}}}
\newcommand{\scp}[2]{\vec{#1}\!\cdot\!\vec{#2}}

\newcommand{\A}{\alpha}
\newcommand{\B}{\beta}
\newcommand{\D}{\delta}
\newcommand{\G}{\gamma}
\newcommand{\Eg}{E_{\mathrm g}}
\newcommand{\ec}{\varepsilon_{\mathrm c}}
\newcommand{\ev}{\varepsilon_{\mathrm v}}
\newcommand{\ve}{\varepsilon}
\renewcommand{\l}{\lambda}
\newcommand{\f}{\phi}
\renewcommand{\j}{\varphi}
\newcommand{\q}{\theta}
\newcommand{\vq}{\vartheta}
\newcommand{\vqp}{\vartheta^+}
\newcommand{\vqm}{\vartheta^-}
\newcommand{\vqpm}{\vartheta^\pm}
\newcommand{\ndr}{\vec{n}\!\cdot\!{\mathrm d}\vec{r}}
\newcommand{\h}{\textbf{H}}
\newcommand{\hEg}{\textbf{H}_{E_{\mathrm g}}}
\newcommand{\hg}{\textbf{H}_{\gamma}}
\newcommand{\hab}{\textbf{H}_{\alpha, \beta}}
\newcommand{\jab}{J_{\A,\B}}
\newcommand{\jg}{J_\gamma}
\newcommand{\jgp}{\widetilde{J}_{\gamma}}
\newcommand{\mos}{MoS$_2$}
\newcommand{\kl}{k_\parallel}
\newcommand{\kp}{k_\perp}
\newcommand{\kpp}{k_\perp^\prime}

\newcommand{\igr}[2][]{\includegraphics[#1]{#2}}


\begin{document}
\title{Boundary conditions for transition-metal dichalcogenide monolayers\\ in the continuum model}

\author{Csaba G. P\'eterfalvi}
\email{csaba.peterfalvi@uni-konstanz.de}
\author{Andor Korm\'anyos}
\author{Guido Burkard}
\affiliation{Department of Physics,
University of Konstanz,
D-78464 Konstanz, Germany}

\date{\today}

\begin{abstract}
We derive the boundary conditions for \mos\ and similar transition-metal dichalcogenide honeycomb (2H polytype) monolayers with the same type of $\scp kp$ Hamiltonian within the continuum model around the K points. In an effective 2-band description, the electron--hole symmetry breaking quadratic terms are also taken into account. We model the effect of the edges with a linear edge constraint method that has been applied previously to graphene. Focusing mainly on zigzag edges, we find that different reconstruction geometries with different edge-atoms can generally be described with one scalar parameter varying between 0 and $2\pi$. We analyze the edge states and their dispersion relation in \mos\ in particular, and we find good agreement with the results of previous density functional theory calculations for various edge types.
\end{abstract}

\maketitle

\section{Introduction}

Semiconducting transition metal dichalcogenides (TMDCs) have attracted much attention lately. This applies especially to their monolayers, which have a direct band gap (\mos, MoSe$_2$, MoTe$_2$, WS$_2$ and WSe$_2$) as opposed to their three-dimensional crystals with an indirect band gap. TMDCs are interesting for a number of applications ranging from flexible and transparent field-effect transistors\cite{lee_flexible_2013,chang_high-performance_2013,das_all_2014,kang_high-performance_2014} to logical\cite{radisavljevic_integrated_2011} and optoelectronic devices.\cite{wang_electronics_2012,britnell_strong_2013,baugher_optoelectronic_2014,pospischil_solar-energy_2014} These applications necessitate the knowledge of the electronic structure of finite flakes and nanoribbons, which in turn also relies on the accurate description of the edges.\cite{bollinger_atomic_2003} Studying the edge states is also very important when charge carriers need to be injected into a layer of TMDC. The edges of TMDCs have already been studied experimentally\cite{lauritsen_size-dependent_2007,zhang_magnetic_2007,tongay_magnetic_2012,van_der_zande_grains_2013,gao_ferromagnetism_2013,yin_edge_2014,zhang_direct_2014} and theoretically\cite{bollinger_atomic_2003,li_mos2_2008,vojvodic_magnetic_2009,ataca_mechanical_2011,erdogan_transport_2012,lu_strain-dependent_2012,pan_tuning_2012,pan_edge-dependent_2012,klinovaja_spintronics_2013,liu_three-band_2013} by several groups. Recently, using STM spectroscopy, experimental evidence of metallic edge states has been found in single layer \mos.\cite{zhang_direct_2014}

\begin{figure}[!b]
  \igr[width=7cm]{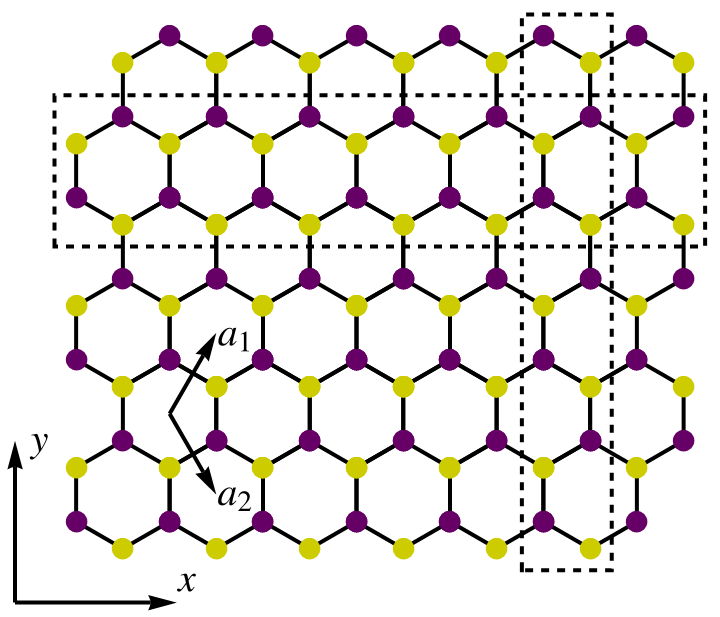}
  \caption{The honeycomb lattice of transition-metal dichalcogenide monolayers. Rectangles with dashed lines show the unit cells of the armchair and zigzag nanoribbons. The width of the nanoribbons can be parametrized by the number $N_\mathrm{d}$ of the A-B dimers in the corresponding honeycomb unit cell without regarding any adatoms attached to the possibly reconstructed edge. In this sense, the width of the framed armchair ribbon would be 14, and that of the framed zigzag ribbon would be 7 in this figure. The actual width of an armchair ribbon is $\frac{1}{2} N_\mathrm{d} a$, and that of a zigzag ribbon is $\frac{\sqrt{3}}{2} N_\mathrm{d} a$, where the lattice constant in \mos\ is $a=|\vec{a_1}|=|\vec{a_2}|=3.1565$\AA.\cite{kormanyos_kp_2015}}
  \label{fig:setup}
\end{figure}

\textit{Ab initio} calculations, like density functional theory (DFT) combined with tight-binding (TB) models\cite{liu_three-band_2013} have allowed many detailed studies of TMDCs. Nevertheless, the description of mesoscopic edge-terminated systems still represents an outstanding challenge for theory. In such cases, a continuum model can offer an efficient way to describe the system. Here, we study the boundary conditions (BCs) and the low-energy in-gap edge states in TMDCs in general, and in \mos\ in particular. Inspired by previous works on the analogous problem in graphene,\cite{mccann_symmetry_2004,akhmerov_boundary_2008} we derive the general boundary conditions for the zigzag and armchair types of edges in monolayer TMDCs, see Fig.~\ref{fig:setup}. The BCs are given in the form of two linear matrix equations, one for the wave function and one for its derivative normal to the edge. These equations can be expressed using an initially unknown matrix $M$, in which the number of free parameters is reduced using the symmetries of the system and the edge. Due to the continuum nature of our method, not all the parameters in $M$ can be determined exclusively by means of symmetry considerations. Some atomistic details of the boundary elude this procedure, and therefore, some parameters need to be fitted to results obtained from other methods with atomistic resolution, such as DFT or density functional based tight-binding (DFTB) methods. We will demonstrate this fitting procedure through the example of zigzag edges in \mos.

\begin{figure}[!t]
  \igr[width=242pt]{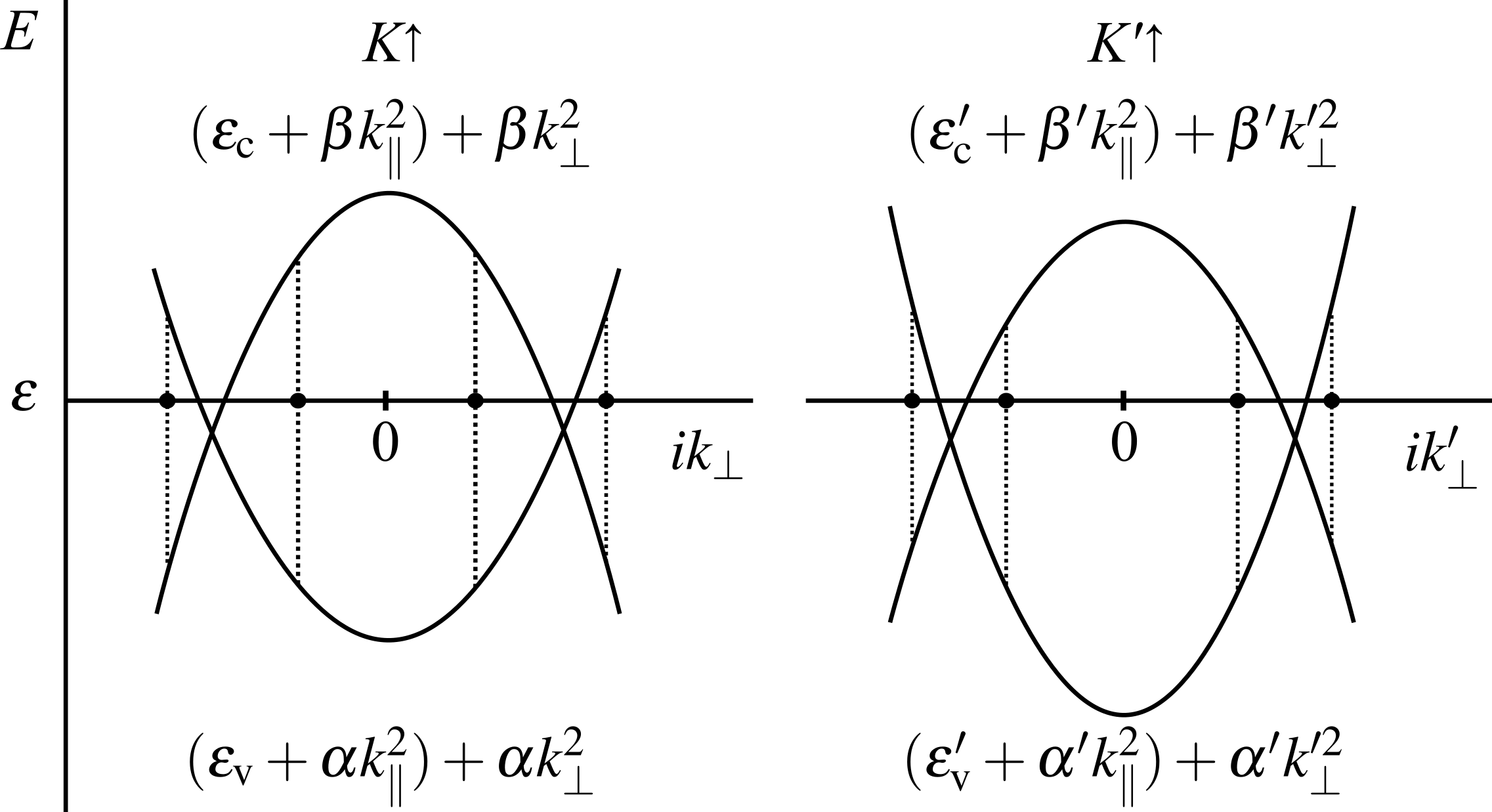}
	\caption{Schematic representation of the eight degenerate states in $k$-space that are mixed by the armchair and zigzag edges. These in-gap states denoted by dots are characterized by the same spin, the same energy $\ve$ and the same quantum number $\kl$, where $\kl$ is the component of $\vec{k}$ parallel to the edge. The dispersion curves are labeled with the corresponding functional forms, which are the diagonal elements of the Hamiltonian $\h$. For solutions localized at the edge, the normal component $\kp$ is imaginary, which results in the opposite curvature of the valence and conduction bands. States from these two bands with the same wave vector compose the corresponding in-gap edge state. From the left to the right, the normal components of the wave vectors are labeled as $-i\kp{}_1$, $-i\kp{}_2$, $i\kp{}_2$, $i\kp{}_1$, $-i\kpp{}_1$, $-i\kpp{}_2$, $i\kpp{}_2$ and $i\kpp{}_1$.}
  \label{fig:eightfold-degeneracy}
\end{figure}

To describe a finite system in \mos, we start with the bulk Hamiltonian for one spin state,
\bnen
\h=\left(\bna{cccc} \A k^2 + \ev & \G k_- & 0 & 0\\ \G k_+ & \B k^2 + \ec & 0 & 0\\ 0 & 0 & \B' k^2 + \ec' & \G k_-\\ 0 & 0 & \G k_+ & \A' k^2 + \ev' \eda\right).
\eden
In this form, the effective $\scp kp$ Hamiltonian\cite{kormanyos_monolayer_2013} is cast into the valley isotropic representation,\cite{akhmerov_boundary_2008} where the blocks of $\h$ correspond to the $K$ and $K'$ valleys, respectively. The basis functions, consisting of mainly the $d$-orbitals of the Mo atoms, are the wave functions of the valence band and the conduction band, as in Ref.~[\onlinecite{kormanyos_monolayer_2013}]. The basis of the 4-spinor thus reads $\{\psi_\mathrm{v},\psi_\mathrm{c},-\psi_\mathrm{c}',\psi_\mathrm{v}'\}$. In the $K^{(\prime)}$ valley, the valence band maximum is $\ev^{(\prime)}$, the conduction band minimum is $\ec^{(\prime)}$, and we also take into account the quadratic terms with $\A^{(\prime)}$ and $\B^{(\prime)}$, which break the electron--hole symmetry. We will refer to the nonequivalence of the two blocks of $\h$, which is due to the spin-orbit interaction (SOI), as \textit{valley asymmetry.} We assume that $\G$ is the same in both valleys, and since we focus on the vicinity of the $K$ points, we neglect the trigonal warping and cubic terms.\cite{kormanyos_monolayer_2013} The distance from the $K$ points is denoted by $k$, and $k_{\pm}=k_x\pm i k_y$. The eigenvalues of the block of $\h$ for the $K$ valley are
\bean
\ve_\pm&=&\frac{1}{2} \left( \ev + \ec + k^2 (\A+\B) \right) \nonumber \\
&\pm& \sqrt{k^2 \G^2 + \frac{1}{4}\left(k^2 (\A-\B) - \Eg \right)^2}\,,
\eean
while $\ve_\pm^\prime(k')$ in the $K'$ valley can be obtained with the respective parameters. The energy gap between the conduction- and valence-band edges is $\Eg^{(\prime)}=\ec^{(\prime)}-\ev^{(\prime)}$. The corresponding eigenstates are
\bean
\left(1,\frac{\ve+\Eg^{}/2-\A k^2}{\G k_-},0,0\right)^\mathrm{T}\ \ \ \mbox{in the $K$, and}\ \ \ \ \ \  \nonumber \\
\left(0,0,1,\frac{\ve-\Eg'/2-\B' k'^2}{\G k'_-}\right)^\mathrm{T}\ \ \ \mbox{in the $K'$ valley,}\ \ 
\eean
where T denotes transposition. At an edge, where only the parallel component of the wave vector $\kl$ remains a good quantum number, the degeneracy is eightfold: twofold for the two valleys, twofold for the two solutions of the wave number at a given $\ve=\ve_+^{(\prime)}(k_1^{(\prime)})=\ve_-^{(\prime)}(k_2^{(\prime)})$ energy, and twofold for the $\pm$ sign of the wave vector's normal component $\kp$, see Fig.~\ref{fig:eightfold-degeneracy}. For now, we neglect the electron spin, and we assume that the armchair and zigzag edges mix these eight states to satisfy the boundary conditions.

The Hamiltonian $\h$ is not specific to \mos, as it can also describe similar transition-metal dichalcogenide monolayers with honeycomb lattices, like MoSe$_2$, MoTe$_2$, WS$_2$ and WSe$_2$.\cite{kormanyos_kp_2015} The $M$ matrix method that we present in this paper is also as general as the Hamiltonian itself. Nevertheless, in the last section, we will demonstrate the use of our method through the example of \mos.\cite{erdogan_transport_2012}

\section{The $M$ matrix method}

Compared to the calculation for graphene,\cite{mccann_symmetry_2004,akhmerov_boundary_2008} an additional boundary condition is needed here for the first derivative of the wave function $\Psi'$, because of the quadratic terms in $\h$. As we will see, the second condition can also be expressed by the same, constant $M$ matrix. Inspired by Ref.~[\onlinecite{mccann_symmetry_2004}], we introduce hard wall boundary conditions for $\Psi$ and $\Psi'$ with the hermitian matrix, $M_0$:
\bnen
\scalebox{1.5}{[}\h + M_0 \D(\scp nr)\scalebox{1.5}{]}\Psi(\vec r)=E\Psi(\vec r)\,,
\label{eq:dirac}
\eden
where \mbox{$\D(\scp nr)$} is the Dirac delta function modeling the confinement potential at the boundary, which crosses the origin here for simplicity, and the unit vector $\vec{n}$ is normal to the boundary and pointing outwards. To obtain information about the unknown matrix $M_0$, we first consider the infinitesimal integral of Eq.~(\ref{eq:dirac}) through the boundary:
\bnen
\lim_{\epsilon\rightarrow0}\int_{-\epsilon}^{+\epsilon}\scalebox{1.5}{[} \hg + \hab + M_0 \D(\scp nr) \scalebox{1.5}{]} \Psi(\vec r)\, \ndr = 0\,, \label{eq:infintfull}
\eden
where the Hamiltonian is split into parts with zeroth order, linear and quadratic terms in $k$ as $\h=\hEg+\hg+\hab$ with indices referring to the parameters they depend on. The integrated zeroth order part and the right hand side of (\ref{eq:dirac}) vanish for any finite wave function. After carrying out the integral in the given limit, we arrive at the following equation for $\Psi=\Psi(\vec r)|_{\scp{n\,}{\,r}=0^-}$ at the boundary:
\bnen
\G \jgp \Psi + A \Psi' + \frac{1}{2} M_0 \Psi = 0, \label{eq:infintfull2}
\eden
where $\Psi'=\Psi'(\vec r)|_{\scp{n\,}{\,r}=0^-}$ denotes the derivative normal to the boundary, the diagonal matrix $A=\mathrm{diag}(\A,\B,\B',\A')$, and $\jgp=\tau_0 \otimes \exp\left(i\frac{\pi}{2}\scp n \sigma \right)$. Here and throughout the paper $\vec s$, $\vec \tau$ and $\vec\sigma$ are denoting vectors of the Pauli matrices acting in the spin space, valley space and the space of the basis functions $\{\psi_\mathrm{v},\psi_\mathrm{c}\}$, with the zeroth component being the two-dimensional identity matrix. Note that the operator $\jgp$ is unitary and antihermitian, and so $\jgp^{\,2}=-1$. Moreover, $A$ combined with the derivative operator is antihermitian. Therefore if we calculate the expectation value of these terms by multiplying them by $\Psi^\dagger$ from the left, the antihermitian operators yield purely imaginary numbers, but the hermitian matrix $M_0$ yields a real number, meaning that they have to be zero separately:
\bean
\G \Psi^\dagger \jgp \Psi + \Psi^\dagger A \Psi' = 0 && \mbox{\ \ and} \label{eq:infintfull3a} \\
\Psi^\dagger M_0 \Psi = 0&&. \label{eq:infintfull3b}
\eean

A second consideration concerns the total current, which can be determined as
\bean
&J&=\frac{i}{\hbar}[\h,\scp nr]= \label{eq:j} \\
&=&\frac{1}{\hbar}\left(\bna{cccc} -2i\A \scp n\nabla & \G e^{-i \chi} & 0 & 0\\ \G e^{i \chi} & -2i\B \scp n\nabla & 0 & 0\\ 0 & 0 & -2i\B' \scp n\nabla & \G e^{-i \chi}\\ 0 & 0 & \G e^{i \chi} & -2i\A' \scp n\nabla \eda\right) \nonumber
\eean
measured in the direction of $\vec{n}=(\cos \chi,\sin \chi)$, which is normal to the boundary, and $\vec{\nabla}$ stands for the gradient operator. We split $J$ in diagonal and off-diagonal parts:
\bnen
J=\jg+\jab=-i\frac{\G}{\hbar} \tau_0 \otimes \exp\left(i\frac{\pi}{2}\scp n \sigma \right) -i \frac{2}{\hbar}A (\scp {n\,} {\,\nabla}),
\eden
where $\jg=-i\frac{\G}{\hbar}\jgp$. It is a natural requirement as a boundary condition that the expectation value of the normal current at the boundary needs to vanish:
\bnen
\langle J \rangle = -i\frac{\G}{\hbar} \left\langle \jgp \right\rangle -i \frac{2}{\hbar} \left\langle A (\scp n \nabla) \right\rangle = 0,
\eden
which can be translated into $\G \Psi^\dagger \jgp \Psi + 2 \Psi^\dagger A \Psi' = 0$ at the boundary. Comparing this to Eq.~(\ref{eq:infintfull3a}), we see immediately that
\bean
\Psi^\dagger \jgp \Psi = 0 && \mbox{\ \ and} \label{eq:vanishingCurrents1}\\
\Psi^\dagger A \Psi' = 0&&, \label{eq:vanishingCurrents2}
\eean
meaning that the expectation values of the diagonal and off-diagonal currents both vanish: $\langle\jab\rangle=\langle\jg\rangle=0$.

We assume now that the boundary problem has at least one solution. From Eq.~(\ref{eq:vanishingCurrents1}), it follows that we could only have four linearly independent solutions in the four-dimensional Hilbert space of $\Psi$, if $\jgp$ was 0, which is not the case. Therefore, the number of the linearly independent solutions for $\Psi$ must be less than four. One could argue that $\Psi'$ should be treated as four additional independent variables adding four more dimensions to the Hilbert space, opening up the possibility to solutions that are linearly independent in the subspace of $\Psi'$, but not in the subspace of $\Psi$. This option can be dismissed if we regard Eq.~(\ref{eq:infintfull2}), which fully determines $\Psi'$ for a given $\Psi$, because $A$ is invertible. This means that the subspace of the solutions in $\Psi$ will determine the corresponding subspace in $\Psi'$ with the same dimensionality.

We now move to the basis spanned by the solutions to $\Psi$, and we denote the size of this subspace by $d$, which is $1\leq d\leq 3$. The first $d$ coordinates of $\Psi$ can thus be chosen arbitrarily and the last $4-d$ have to be zero. For Eq.~(\ref{eq:vanishingCurrents1}), this means that the top left $d \times d$ block of $\jgp$ must be 0 in this basis. At the same time, the last $4-d$ columns of $M_0$ can be chosen freely. We then define the matrix $M_1$ in the following way:
\bnen
\def\arraystretch{1.5}
M_1=\left(\begin{array}{c|c}
0 & -Y^\dagger \\
\cline{1-2}
-Y & 0
\end{array}\right), \label{eq:M1}
\eden
where $Y$ is a $(4-d) \times d$-sized matrix containing the matrix elements of the bottom left block of $\jgp$ (row indices from $d+1$ to 4 and column indices from 1 to $d$). The top left $d \times d$-sized block contains zeros, and the same holds for the bottom right $(4-d) \times (4-d)$-sized block. Since $\jgp$ is unitary and antihermitian, it is easy to prove that it consists of the same blocks as $M_1$, except for the sign change of the bottom left block. It follows from here that
\bnen
\jgp\Psi+M_1\Psi=0 \label{eq:infintlin}
\eden
for every solution of the boundary problem, and $M_1$ is a hermitian matrix. Eq.~(\ref{eq:infintlin}) multiplied by $\G$ and subtracted from Eq.~(\ref{eq:infintfull2}) leads us to
\bnen
A \Psi' + M_2 \Psi = 0, \label{eq:infintquad}
\eden
where we introduced $M_2=M_0/2-\G M_1$, which must be also hermitian, given that both $M_0$ and $M_1$ have this property.

Using that $\jgp^{\,2}=-1$, Eq.~(\ref{eq:infintlin}) can be rewritten as $\jgp M_1 \Psi = \Psi$, and this boundary condition becomes
\bnen
M \Psi = \Psi\label{eq:bc1}
\eden
with the definition $M \equiv \jgp M_1$. It follows from the similar block-structure of $\jgp$ and $M_1$, and from $Y^\dagger Y=1$, that $M$ is a diagonal matrix, the first $d$ elements of which are 1 along the diagonal, and the last $4-d$ are $-1$. From here, it is straightforward to see that $\{\jgp,M\}=0$, $M$ is hermitian and unitary, and $M_1$ must be unitary too. From $\{\jgp,M\}=0$, the properties of $\jgp$ and the basis-invariance of the trace, it also follows that $M$ is traceless, therefore it has two eigenvalues of 1 and two of $-1$, making $d=2$, and all the blocks in~(\ref{eq:M1}) $2\times 2$. Another way to show this can be found in the basis where $\jgp$ is a real antisymmetric matrix, and $M_1$ is consequently a real symmetric matrix. The trace of the product of two such matrices is always 0. At this point, we see that $M$ has two eigenvectors with eigenvalues of 1, and two other eigenvectors with eigenvalues of $-1$. Note that the basis consisting of these eigenvectors is not fully specified as long as these pairs of eigenvectors can be rotated in their respective two-dimensional subspaces. We can specify the basis by fixing $Y$. For example, if $Y=\sigma_0$, then $\jgp=-i \tau_2 \otimes \sigma_0$ and $M_1=- \tau_1 \otimes \sigma_0$. Independently from the specific choice of $Y$, we can call this basis the eigenbasis of $M$, because $M=\tau_3 \otimes \sigma_0$ is diagonal. The subspace of $M$ with eigenvalues $\pm1$ is mapped by $\jgp$ into the subspace with eigenvalues $\mp1$, which is a basis-invariant feature of course.

From equations~(\ref{eq:vanishingCurrents2}) and~(\ref{eq:infintquad}), we obtain that $\Psi^\dagger M_2 \Psi=0$, which means that the top left $2 \times 2$ block of the hermitian matrix $M_2$ must be 0 in the eigenbasis of $M$. In this basis, one can easily show that \mbox{$(M+1)M_2(M+1)=0$,} which also holds however in every other basis. From here, we obtain $(M+1)M_2 \Psi=0$, which combines with~(\ref{eq:infintquad}) to the boundary condition for the derivative of the 4-spinor:
\bnen
(M+1)A \Psi'=0.\label{eq:bc2}
\eden
Note that if $A=0$, then (\ref{eq:bc2}) is trivial, and we formally return to the solution for gapped graphene with the only boundary condition of Eq.~(\ref{eq:bc1}) to be solved with four linearly independent functions at the edge.

The two boundary conditions for $\Psi$ and $\Psi'$, Eqs.~(\ref{eq:bc1}) and~(\ref{eq:bc2}) represent the first main results of this work. Both equations are expressed with the same $M$ matrix, which is hermitian, unitary, and satisfies $\{\jgp,M\}=0$. These are the same conditions that led to Eq.~(2.8) of Ref.~[\onlinecite{akhmerov_boundary_2008}]:
\bnen
M=\sin\l\, \tau_0 \otimes (\scp \xi \sigma) + \cos\l\, (\scp \nu \tau) \otimes (\scp \mu \sigma),\label{eq:ML}
\eden
where $\vec\xi$, $\vec\nu$ and $\vec\mu$ are three-dimensional unit vectors, such that $\vec\xi$ and $\vec{\mu}$ are orthogonal to each other and also to $\vec{n}_{\mathrm b}$, which is in turn normal to the boundary, and pointing outwards. The vector $\vec\nu$ can be parametrized by the polar angle $\j$ and the azimuthal angle $\f$, and the vectors $\vec\mu$ and $\vec\xi$ with a single polar angle $\q$ and $\q+\pi/2$. The optional minus sign in the latter, can be merged with the sign of the angle $\l$. Assuming that no external magnetic field is present, we now invoke time reversal symmetry. In order to explicitly take into account the spin degree of freedom, we extend the Hilbert space. The Hamiltonian $\h$ must be extended with the same two blocks along the diagonal, but this time with $\ec'$, $\ev'$, $\A'$, and $\B'$ parametrizing the $K$ valley, and $\ec$, $\ev$, $\A$, and $\B$ the $K'$ valley. The $2 \times 2$ blocks of $\h$ can now be labeled as
\bnen
\left[\bna{cccc} K\!\uparrow & & &\\ & K'\!\uparrow & &\\ & & K\!\downarrow &\\ & & & K'\!\downarrow\eda\right].\label{eq:full-basis}
\eden
This $8\times 8$ Hamiltonian is commuting with the time reversal operator
\bnen
T=-s_x \otimes \tau_y \otimes \sigma_y \, \mathcal{C},
\eden
where $\mathcal{C}$ is the operator for complex conjugation. Upon time reversal, the real components of the $\vec k$ vector change sign, thus also changing the valley, and the electron spin flips. The edges on the other hand, do not couple different electron spin states, so in the basis defined at (\ref{eq:full-basis}), two blocks of $M$ appear along the diagonal possibly mixing states in different valleys but only with the same spin. Allowing that the parameter angles defined at (\ref{eq:ML}) are different for the two spin states, we require that $\mathrm{diag}(M(\l,\j,\f,\q),M(\l',\j',\f',\q'))$ commutes with $T$, where diag denotes a (block-)diagonal matrix with the arguments along the diagonal. This can be solved with $\l'=-\l$, and $\j'=\j$, $\f'=\f$, and $\q'=\q$, and unless $\A+\B=\A'+\B'=0$, electron--hole symmetry or valley symmetry cannot be used to further reduce the number of the four free parameters of $M$, unlike in Ref.~[\onlinecite{akhmerov_boundary_2008}].

There is another symmetry of $\h$ that swaps the valleys and flips the spin, but does not change the sign of the momentum relative to the $K$ points. This is represented by the operator
\bean
Q&=&-s_x \otimes \tau_y \otimes \sigma_y \times\\
&&\times\mathrm{diag}(-k_+,k_-,-k_+,k_-,-k_+,k_-,-k_+,k_-)/k.\nonumber
\eean
This symmetry corresponds to a $\pi$-rotation with respect to the $y$ axis at an armchair edge (see Fig.~\ref{fig:setup}) and results in spin-degeneracy there, in contrast to zigzag edges, where $Q$ represents no symmetry and we observe spin-splitting. Unfortunately, this symmetry operation cannot be used to gain further information on $M$ by requiring $[M,Q]=0$, because of the momentum-operators in $Q$ acting on edge states, which are superpositions of the bulk solutions with different $\kp$.

\begin{figure}[!t]
  \igr[width=242pt]{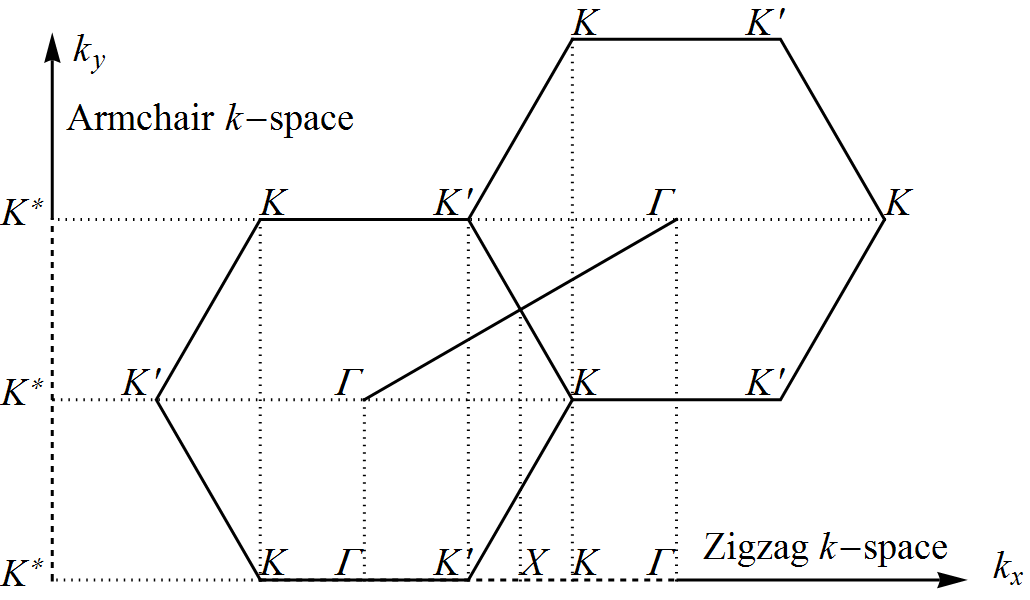}
  \caption{Projections to the reciprocal spaces of zigzag and armchair nanoribbons. In armchair ribbons the valleys are fully mixed and all the $K$ points are equivalent, while in zigzag ribbons, the $K$ and $K'$ points remain independent, the valley mixing is zero.}
  \label{fig:kspace}
\end{figure}

Apart from the mass term, and the valley-dependent quadratic terms $\A$ and $\B$, there are more differences between graphene and the transition-metal dichalcogenide monolayers which we focus on here. While in graphene the basis functions are the orbitals in the two sublattices, the basis we are using here consists of the extended wave functions of the valence and conduction bands. This makes it less straightforward to see how the angles $\l$, $\j$, $\f$ and $\q$ relate to the geometry of a given edge. We will discuss here the two special cases of zigzag and armchair edges (Fig.~\ref{fig:setup}~and~\ref{fig:kspace}), with a strong emphasis on the zigzag edge for being energetically favorable\cite{van_der_zande_grains_2013,lauritsen_size-dependent_2007} and more relevant from the applications' point of view. From Fig.~\ref{fig:kspace}, it is clear that an armchair edge mixes the $K$ and $K'$ valleys, while these remain non-equivalent in the 1D Brillouin zone of a zigzag edge. This suggests that the polar angle $\j$ responsible for the valley mixing must be 0 or $\pi$ for zigzag, and from symmetry considerations, we also deduce that $\j=\pi/2$ causing the maximal mixing of the valleys, must belong to the armchair orientation, just like in graphene.\footnote{Note that a finite $\l$ parameter in $M$ does not allow complete mixing of the valleys at an armchair edge. Only at vanishing valley asymmetry ($\l=0$) can the valley mixing be complete with $\j_\mathrm{armchair}=\pi/2$.} This also means that $\f$ drops out of the equations~(\ref{eq:bc1}) and~(\ref{eq:bc2}) for zigzag, leaving only $\l$ and $\q$ to characterize the various zigzag edges. As it turns out, $\f$ matters for armchair edges, but only if edge states from different edges can interfere. For a single armchair edge, it is also $\l$ and $\q$ that determine the edge states and their dispersion relation. Without studying the atomistic details of an edge, based on symmetry considerations alone the values of $\l$ and $\q$ cannot be inferred within the continuum model. However, fitting to \textit{ab initio} band structures is a straightforward way to relate the different edge geometries to different values of these parameters.

With $\j=0$ for zigzag, $M$ can be simplified to
\bnen
M_\mathrm{zz}=\left(\bna{cccc} \cos(\vqp) & \sin(\vqp) &0&0\\
\sin(\vqp) & -\cos(\vqp) &0&0\\
0&0& -\cos(\vqm) & -\sin(\vqm) \\
0&0& -\sin(\vqm) & \cos(\vqm) \eda\right),
\eden
where $\vqpm=\q\pm \l$. The $\j=\pi$ case is already included if we shift $\vq$ by $\pi$. Since the zigzag edge does not mix the valleys, $\vqp$ and $\vqm$ can be fitted separately. As we know from above, the same $M_\mathrm{zz}$ also works for the opposite spin, but with $-\l$. Note that $\l$ is expected to be small, for it is finite only because of the valley asymmetry. If all the parameters in $\h$ were the same in both valleys, then $\h$ would commute with $-\tau_y \otimes \sigma_y \, \mathcal{C}$ (the $4\times 4$ off-diagonal block of $T$), which in turn has to commute also with $M_\mathrm{zz}$, resulting in $\l=0$ as in graphene (see Eq.~(2.10) of Ref.~[\onlinecite{akhmerov_boundary_2008}]).

\section{Single edges and nanoribbons}

\subsection{General discussion}

Here we apply the $M$ matrix method to various kinds of edges. Since we are interested in the in-gap edge states, the normal component of the wave vector $\kp$ will always be imaginary. Out of the eight edge-solutions of the eigenvalue problem, we can only work with the four that are decaying away from the edge of a half-infinite system and we discard the diverging solutions. This leads to four unknown coefficients and eight equations defined by~(\ref{eq:bc1}) and~(\ref{eq:bc2}). However, the rank of each of the $4\times 4$ matrices defining this linear system of equations is 2, yielding in total four equations to solve for the four coefficients. The determinant of this $4\times 4$ matrix is zero for the non-trivial solutions; this delivers the dispersion relation of the edge states, and the null-space of the matrix provides us with the coefficients.

In a nanoribbon, the dispersion relation can be calculated in the same way, but with eight states resulting in a determinant of an $8\times 8$ matrix. Since the edge states are concentrated in the vicinity of the edges, in a few nm wide nanoribbon, they become independent. Further increments of the width do not affect the bands of the edge states any more, and the bands of the single edges independently add up to form the band structure of the nanoribbon. It is important to note that two edges, characterized by $\l_{1,2}$ and $\q_{1,2}$ (or by $\vqpm_{1,2}$ in zigzag), can be combined to form a nanoribbon, if we attribute $-\l$ and $-\q$ ($-\vqpm$) to the edge bordering the ribbon on the opposite side relative to the one that was used to obtain these angles. Eg.~in the case of an armchair ribbon with equivalent edges, $\l$, $\q$, $-\l$ and $-\q$ has to enter the above mentioned $8\times 8$ matrix to give rise to the correct dispersion relation originating from the combined effect of the two identical edges.

This method can be generalized to flakes with more complex shapes and mixed types of edges, like equilateral triangles, rectangles, hexagons, etc. Each edge has its respective $\q$ and $\l$, and by studying the symmetries of the system of equations posed by the boundary conditions, it must be determined how to couple these to form the boundaries of the same flake. In the case of armchair edges, the relative distance and orientation of the edges also matter if the corresponding states are overlapping, meaning that the knowledge of the angles $\f$ is also required for the accurate description. As we see at the end of this section through a simple example in graphene, this can be determined from an appropriate TB model.

\subsection{Application to \mos}

To demonstrate the use of our method, we calculate the edge states in \mos. To this end, we need the material parameters appearing in the Hamiltonian. Because of the mirror symmetry of the lattice to the $x-y$ plane, $\h$ is block-diagonal in the $z$-component of the real spin and takes the form of (\ref{eq:full-basis}) with the valley-dependent sets of parameters \{$\A$, $\B$, $\G$, $\ev$, $\ec$\} and \{$\A'$, $\B'$, $\G$, $\ev'$, $\ec'$\}. We assume that no external field or special edge geometry breaks this symmetry of the bulk, and therefore the full Hamiltonian including $M$ also preserves this symmetry.
Our continuum model, which describes the edge states as linear combinations of the bulk states using only the basis of the valence and the conduction bands, will not be able to describe all the in-gap states. For example, states that are almost entirely localized on the terminating row of sulfur atoms (see eg. Fig.~5 of Ref.~[\onlinecite{bollinger_atomic_2003}]) cannot be expanded in the basis we use. The bulk wave functions of this basis require that the parameters appearing in $\h$ to be extracted from a bulk DFT study. The values of these bulk parameters are taken from Ref.~[\onlinecite{kormanyos_kp_2015}], where $\G$ is reported to be spin-independent: $\A=-2.16$~eV\AA${}^2$, $\A'=-2.62$~eV\AA${}^2$, $\B=4.09$~eV\AA${}^2$, $\B'=4.35$~eV\AA${}^2$, $\G=2.76$~eV\AA, $\Eg=\ec-\ev=1.821$~eV, $\Eg^\prime=\ec'-\ev'=1.67$~eV, $\ec-\ec'=3$~meV and $\ev'-\ev=148$~meV. Note that the contribution of the free Hamiltonian $\hbar^2/(2m_{\rm e})=3.81$~eV\AA${}^2$ is included in the $\A$, $\A'$, $\B$ and $\B'$ parameters, where $m_{\rm e}$ is the free electron mass.

In what follows, we fit and compare our results to the DFTB study of Erdogan \textit{et al.}~for nanoribbons.\cite{erdogan_transport_2012} Although SOI was not considered, the dispersion curves of sufficiently wide (32.8~\AA) nanoribbons with several different kinds of edges were calculated, which offers the opportunity to fit $\vq$ edge-by-edge assuming that the two sides are independent. Such DFT calculations are typically done with nanoribbons and not with single edges because of the periodic boundary condition. However, we can consider the edge bands independent in nanoribbons over a few nm wide, because exponentially decaying edge states on either sides become negligible at the opposite edge and they do not interfere.
Here we first adopt the assumption that the DFTB band structure is actually spin-resolved and belongs to the spin-up states in the $K'$ valley, and then we repeat the calculation with the assumption that the same bands belong to spin-down states in the same valley. With this fitting to spin-independent bands, we aim to demonstrate the use of our method, rather than providing quantitatively accurate results. Accuracy could be obtained from fitting to non-magnetic, spin-resolved bands, but to our best knowledge, no such study has been published to this day.

\begin{figure*}[!t]
    \captionsetup[subfigure]{labelformat=parens}
    \subfloat[\ Antisymmetric edges\hfill\ ]
    {\label(SubFig1){fig:bs-antisymmetric}
    \igr[width=180pt,trim=0 -45 -200 0]{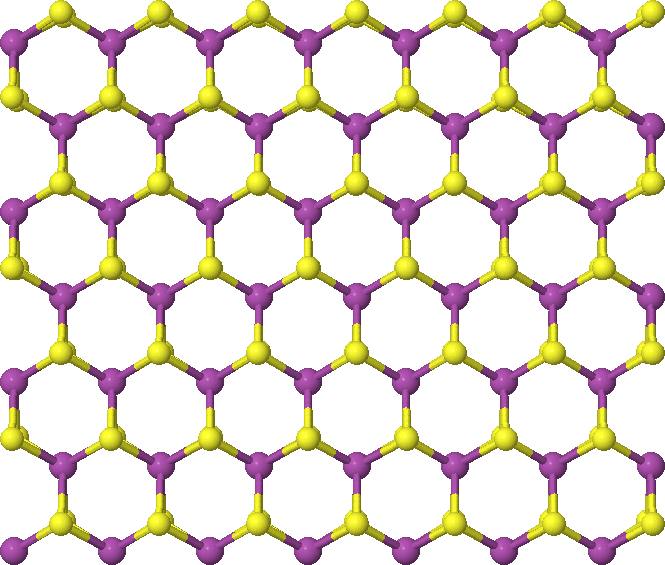}}
    \subfloat{\igr[width=45pt,trim=0 -45 -100 0]{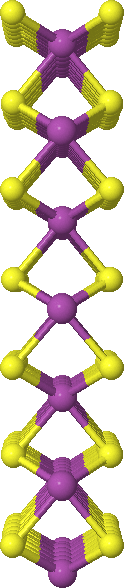}}
    \captionsetup[subfigure]{labelformat=empty}
    \subfloat[\hspace{50pt} $\vqm_1=1.795$, $\vqm_2=\pi+1.762$.\hfill\ ]{\igr[height=130pt,trim=-100 30 -100 0]{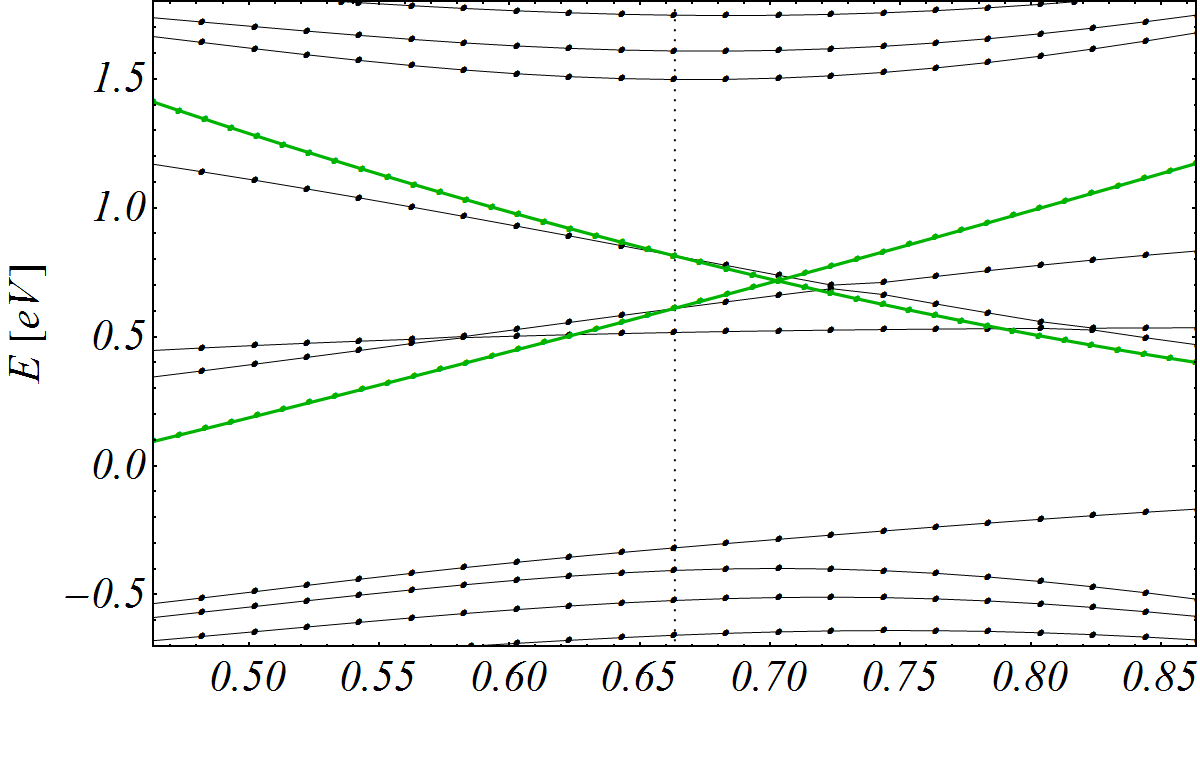}}
    \addtocounter{subfigure}{-2}
    \captionsetup[subfigure]{labelformat=parens}
    \subfloat[\ S${}_2$-stripe edges\hfill\ ]
    {\label(SubFig2){fig:bs-S2-stripe}
    \igr[width=180pt,trim=0 -30 -200 0]{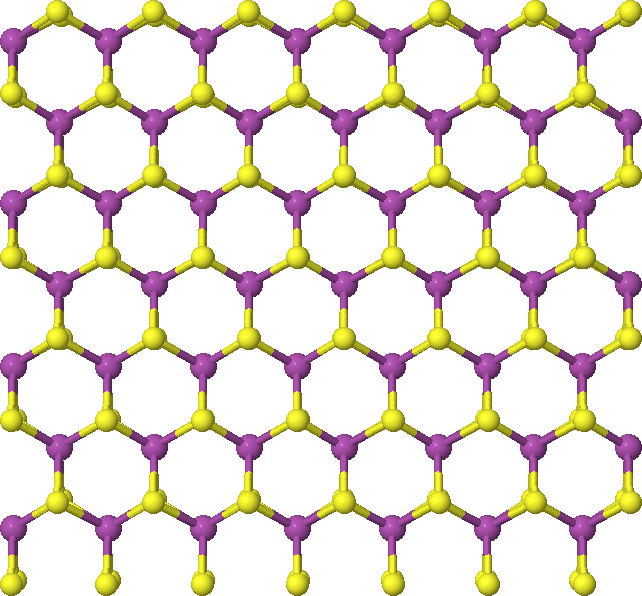}}
    \subfloat{\igr[width=45pt,trim=0 -30 -100 0]{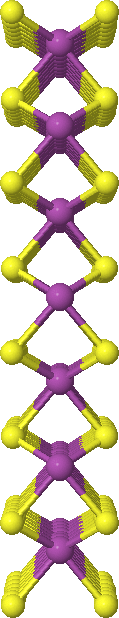}}
    \captionsetup[subfigure]{labelformat=empty}
    \subfloat[\hspace{50pt} $\vqm_1=1.768$, $\vqm_2=\pi+1.840$.\hfill\ ]{\igr[height=130pt,trim=-100 30 -100 0]{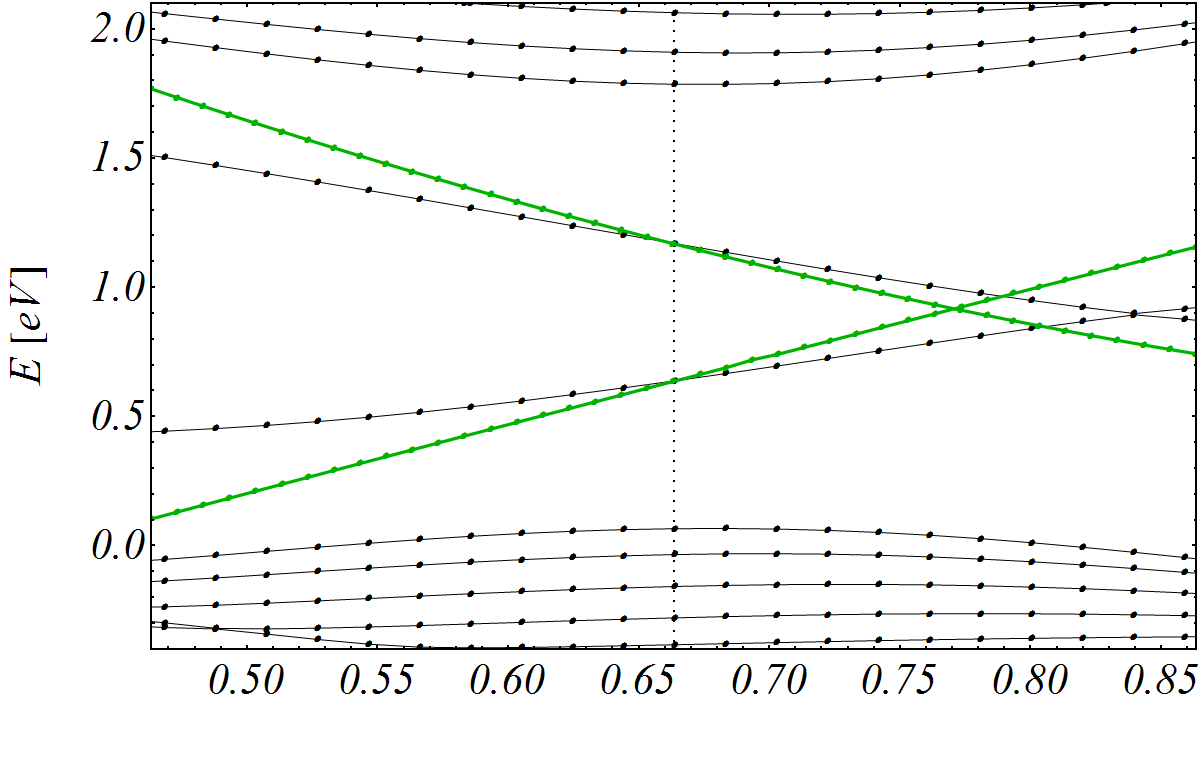}}
    \addtocounter{subfigure}{-2}
    \captionsetup[subfigure]{labelformat=parens}
    \subfloat[\ S-dimer edges\hfill\ ]
    {\label(SubFig3){fig:bs-S-dimer}
    \igr[width=180pt,trim=0 -30 -200 0]{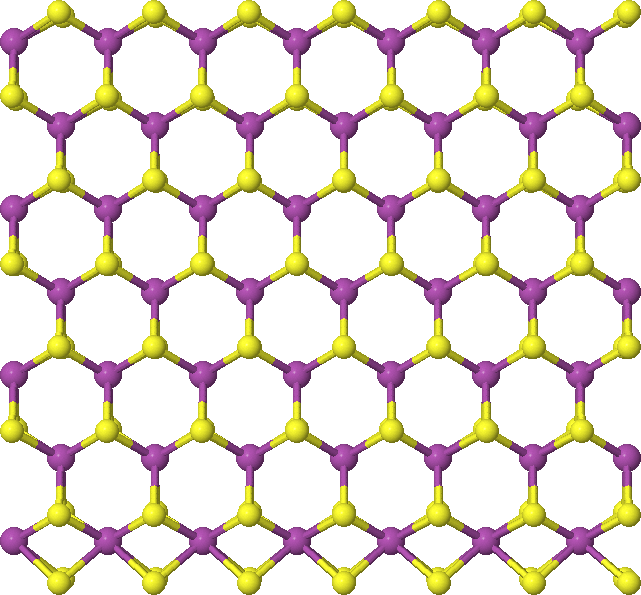}}
    \subfloat{\igr[width=45pt,trim=0 -30 -100 0]{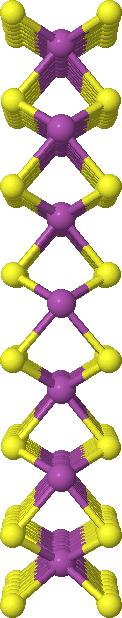}}
    \captionsetup[subfigure]{labelformat=empty}
    \subfloat[\hspace{50pt} $\vqm_1=1.791$, $\vqm_2=\pi+1.810$.\hfill\ ]{\igr[height=130pt,trim=-100 30 -100 0]{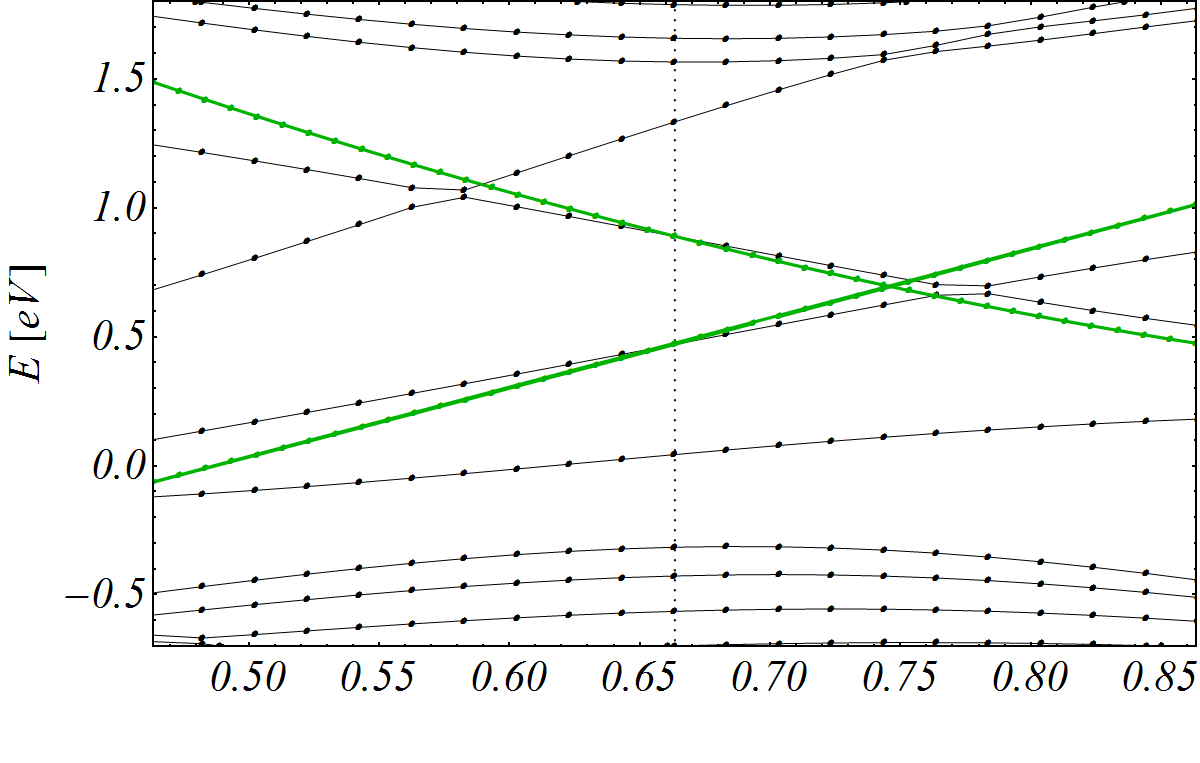}}
    \addtocounter{subfigure}{-2}
    \captionsetup[subfigure]{labelformat=parens}
    \subfloat[\ S-half edges\hfill\ ]
    {\label(SubFig4){fig:bs-S-half}
    \igr[width=180pt,trim=0 -30 -200 0]{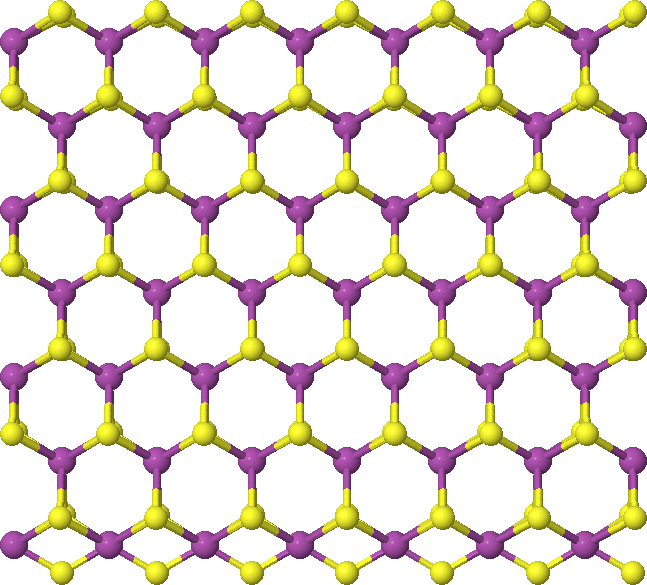}}
    \subfloat{\igr[width=45pt,trim=0 -30 -100 0]{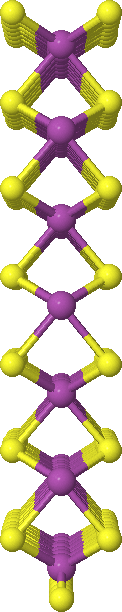}}
    \captionsetup[subfigure]{labelformat=empty}
    \subfloat[\hspace{50pt} $\vqm_1=1.780$.\hfill\ ]{\igr[height=130pt,trim=-100 30 -100 0]{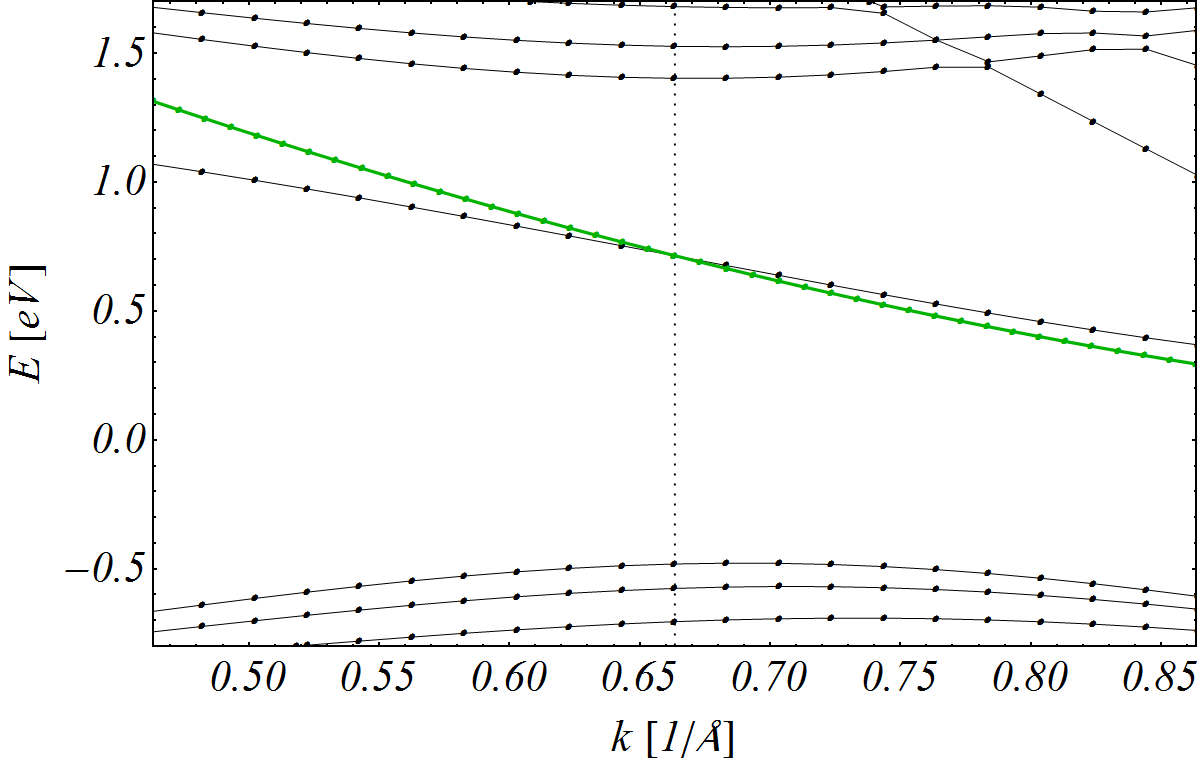}}
    \caption{On the left, various 12 dimers (32.8~\AA) wide zigzag nanoribbons from Ref.~[\onlinecite{erdogan_transport_2012}], with the corresponding band structures on the right. The edge state bands from the continuum model for spin-up states are denoted by thick, green (gray) lines, and those from the DFTB calculation from Ref.~[\onlinecite{erdogan_transport_2012}] are denoted by thin, black lines. The dotted lines show the $K'$ point, and the wave number is measured from the $\Gamma$ point in these figures. The thick green (gray) curves correspond to the states at the two edges fitted with $\vqm_1$ and $\vqm_2$, where $\j_1=\j_2=0$. The dots represent the data points the curves were fitted to.}
		\label{fig:bs}
\end{figure*}

\begin{figure}[!t]
  \igr[width=242pt,trim=0 70 0 60,clip]{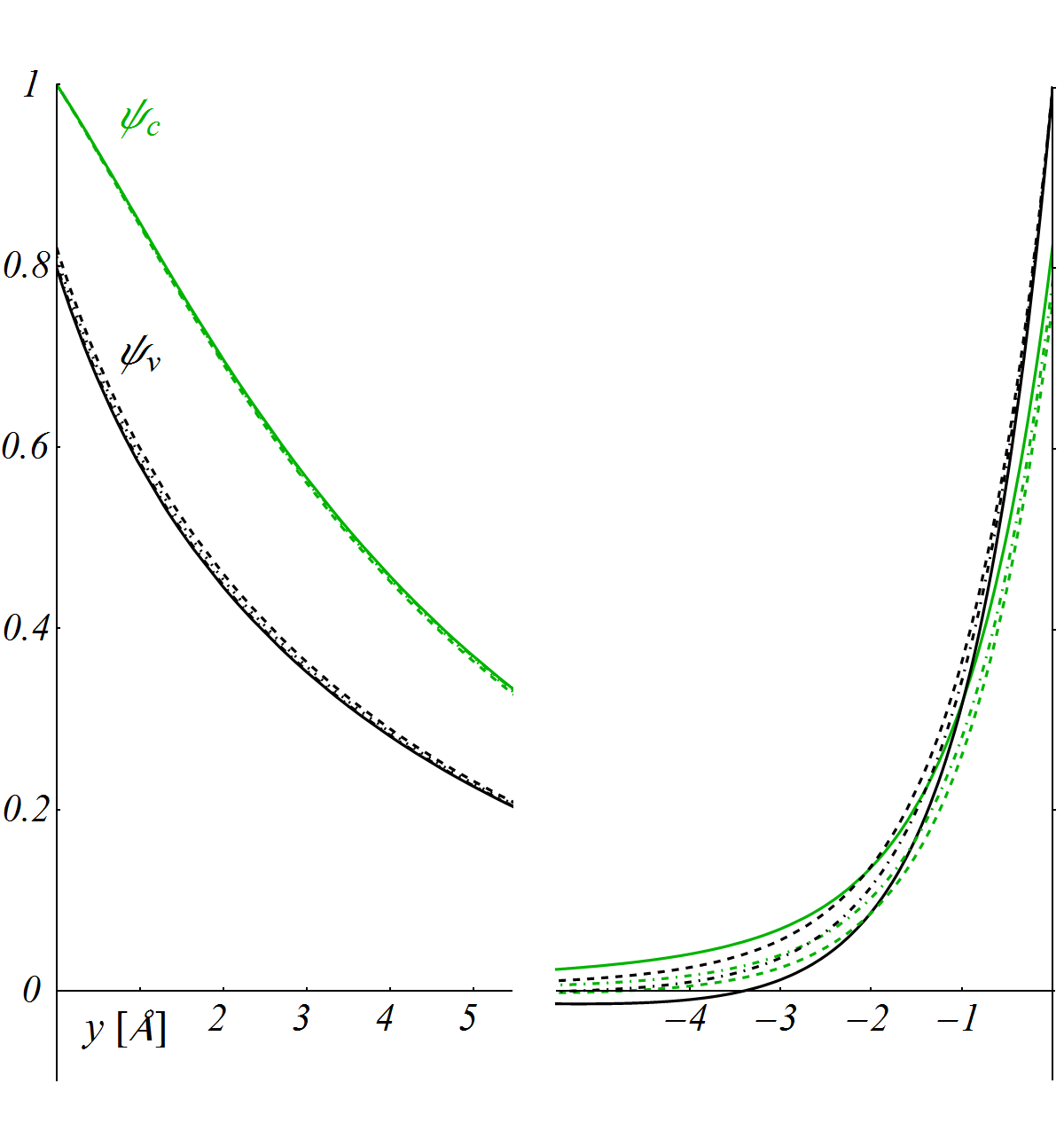}
  \caption{Coefficients of $\psi_\mathrm{v}$ and $\psi_\mathrm{c}$, \textit{ie.}~the wave function at $k_x=0$ in the $K\!\!\downarrow$ and $K'\!\!\uparrow$ valleys at the zigzag edges, within the width of two dimers $\sqrt{3} a$. Solid lines describe the antisymmetric edges (Fig.~\ref{fig:bs-antisymmetric}), dashed lines the S${}_2$-stripe edges (Fig.~\ref{fig:bs-S2-stripe}) and dashed-dotted lines the S-dimer edges (Fig.~\ref{fig:bs-S-dimer}). Dotted lines appearing only on the left describe the S-half ribbon, which does not support edge states on the right (Fig.~\ref{fig:bs-S-half}). The wave function is calculated for independent edges with the $\vq$ values from Fig.~\ref{fig:bs}. Note that all the components of $\Psi$ are real, and the maximum of $\psi_\mathrm{c}$ ($\psi_\mathrm{v}$) on the left (right) is always normalized to~1. For the corresponding energies, see the band structures in Fig.~\ref{fig:bs}~at the $K'$ point.}
  \label{fig:zz-wfcn}
\end{figure}

The fitted spin-up bands are plotted in Fig.~\ref{fig:bs}, together with the original DFTB bands and nanoribbon structures from Ref.~[\onlinecite{erdogan_transport_2012}].\footnote{The spin-down bands can be fitted equally well: $\vqp_1=1.762$, $\vqp_2=\pi+1.817$ for the antisymmetric edges, $\vqp_1=1.737$, $\vqp_2=\pi+1.886$ for the S${}_2$-stripe edges, $\vqp_1=1.758$, $\vqp_2=\pi+1.860$ for the S-dimer edges, and $\vqp_1=1.747$ for the S-half edges.} To fix the energy offset of the edge bands, we match the center of the bulk gap to the center of the nanoribbon energy gap by a rigid energy shift for the whole band structure. Then we adjust the energy of our edge band at the $K'$ point to that of the DFT-calculated band by tuning the only free parameter $\vq$ for each single edge band separately. The fitted $\vq$ values are unique apart from the $2\pi$ periodicity, and these were used for the plots of Fig.~\ref{fig:zz-wfcn}~as well.\footnote{At the $K'$ point, +0.01 change in $\vqpm_1$ means approximately $10$~meV increase in energy and the same change in $\vqpm_2$ results in roughly 45~meV decrease in energy. Nevertheless, all the angles $\vqpm_{1,2}$ were fitted with an accuracy of 0.001.}
It is apparent from Fig.~\ref{fig:bs} that all four band structures share an in-gap state with a negative slope. This suggests that this band belongs to the same edge that all the four nanoribbons have in common, namely the purely sulfur terminated zigzag edge (the top edges of the ribbons in Fig.~\ref{fig:bs}). To this band, we can fit the angle $\vqm_1$ with almost the same value in every cases. The edges on the other side are described by $\vqm_2$, and in this case, we can only fit to one band without ambiguity even though in two cases (Fig.~\ref{fig:bs}(a) and (c)), other in-gap bands appear in DFT, which are missing from the continuum model. Note that the edge with one S atom per Mo supports no edge state at all, see Fig.~\ref{fig:bs-S-half}.

The deviations from the DFTB band structure can be explained with the following factors: (i) in the DFTB calculation, the spin-orbit coupling was neglected; (ii) the input parameters for our calculation ($\A$, $\B$, $\G$ and $\Eg$) were extracted from a bulk DFT calculation\cite{kormanyos_kp_2015} with different exchange-correlation potential, etc; and finally (iii) we used an effective, 2-band Hamiltonian neglecting the trigonal warping and the cubic terms.\cite{kormanyos_monolayer_2013} Regarding all these differences, the agreement between the band structures is reasonably good. The most significant deviation manifests itself in the larger steepness of the bands that have lower energy at the $K'$ point. Note however that this band is moderately steeper also in Refs.~[\onlinecite{gibertini_emergence_2015}] relative to our reference,\cite{erdogan_transport_2012} and the derivative of our bands agrees especially well with the tight binding results in Fig.~9 from Ref.~[\onlinecite{liu_three-band_2013}], even if we need to use slightly different $\vq$ angles to fit our bands to theirs.

The corresponding components of the wave function can be seen in Fig.~\ref{fig:zz-wfcn}, where the real-valued coefficients of $\psi_\mathrm{v}$ and $\psi_\mathrm{c}$ are plotted along the dimension normal to the various kinds of zigzag edges in Fig.~\ref{fig:bs}.
Here, the left edge always has the same, purely sulfur terminated structure, whereas the right edge has different reconstruction in the four cases. Since the S-half ribbon has no edge states on the right (see Fig.~\ref{fig:bs-S-half}), the dotted lines only appear on the left. The differences in the wave function on the right manifest in slightly different $\vq$ values for the otherwise identical left edges (see Fig.~\ref{fig:bs}), and here, this prevents the $\psi_\mathrm{v}$ curves and also the $\psi_\mathrm{c}$ curves from exact overlap. This is due to the small, yet finite contribution of the electrons from the opposite edge 32.8~\AA\ away. On the left, the overall predominance of the conduction band can be seen, but on the right, the contribution of the two bands is roughly the same. Accordingly, the left edge state will always have the higher energy, see Fig.~\ref{fig:bs}. Using Eq.~(\ref{eq:vanishingCurrents2}), we can verify a direct relation between these curves, namely that \mbox{$\psi_\mathrm{v} \A' \psi_\mathrm{v}^\prime + \psi_\mathrm{c} \B' \psi_\mathrm{c}^\prime = 0$} holds at both edges. The ratio of the $\A'$ and $\B'$ parameters explains the relative height and steepness of the $\psi_\mathrm{v}$ and $\psi_\mathrm{c}$ curves at the two ends.

Note that in graphene, there are basically only two kinds of zigzag edges, and even they are related by symmetry. One edge is always terminated by atoms of sublattice A, and the other one by atoms of sublattice B. In our case, these atoms are different elements, we lose inversion symmetry, and also the fact that molybdenum atoms form bonds with six out-of-plane sulfur atoms increases the number of possible edge terminations. In the case of the antisymmetric edges (Fig.~\ref{fig:bs-antisymmetric}), the purely S and purely Mo terminated edges correspond to the solid curves in Fig.~\ref{fig:zz-wfcn}~on the left and on the right respectively. This S--Mo antisymmetry constitutes the major difference between the wave functions on the two sides. The dangling bonds of the Mo atoms offer further possibility for various reconstructions of the edge with additional S atoms thus modifying the picture to some lesser extent on the right, but it does not change qualitatively. As we see, the obtained values for the $\vqm$ angles on the right ($\vqm_2$) and on the left ($\vqm_1$) are also forming two distinct groups, emphasizing that the sublattice asymmetry is more important than the effect of the various edge reconstructions on the Mo-terminated side.
This is also underlined regarding the energy of these states: the maximal energy difference at $k_x=0$ between the two sides is 531~meV and the slopes of the bands differ in sign. The maximal energy difference between the various reconstructions on the right relative to the middle of the gap is 351~meV, and the slopes of the bands are almost the same. And finally, the maximal difference on the left due to edge states on the right reaching the left edge is only 26~meV. Remember that in the context of the continuum model, the different values of $\vqm_1$ can be understood to be due to the small, yet finite contribution of the states from the opposite edges at a distance of 32.8~\AA. Otherwise, one kind of edge is characterized by one pair of angles $\vqpm$, which in the case of the purely S terminated left edge would be around $1.78$. The precise determination of $\vqpm_1$ would necessitate fitting to \textit{ab initio} calculations with somewhat wider ribbons, or the simultaneous fitting of $\vqpm_1$ and $\vqpm_2$ in a ribbon explicitly considering the interference of the edge states. This accuracy however would be justified only if we were fitting to a spin-resolved band structure accounting for the spin-orbit interaction as well.

\begin{figure}[!t]
  \igr[width=242pt]{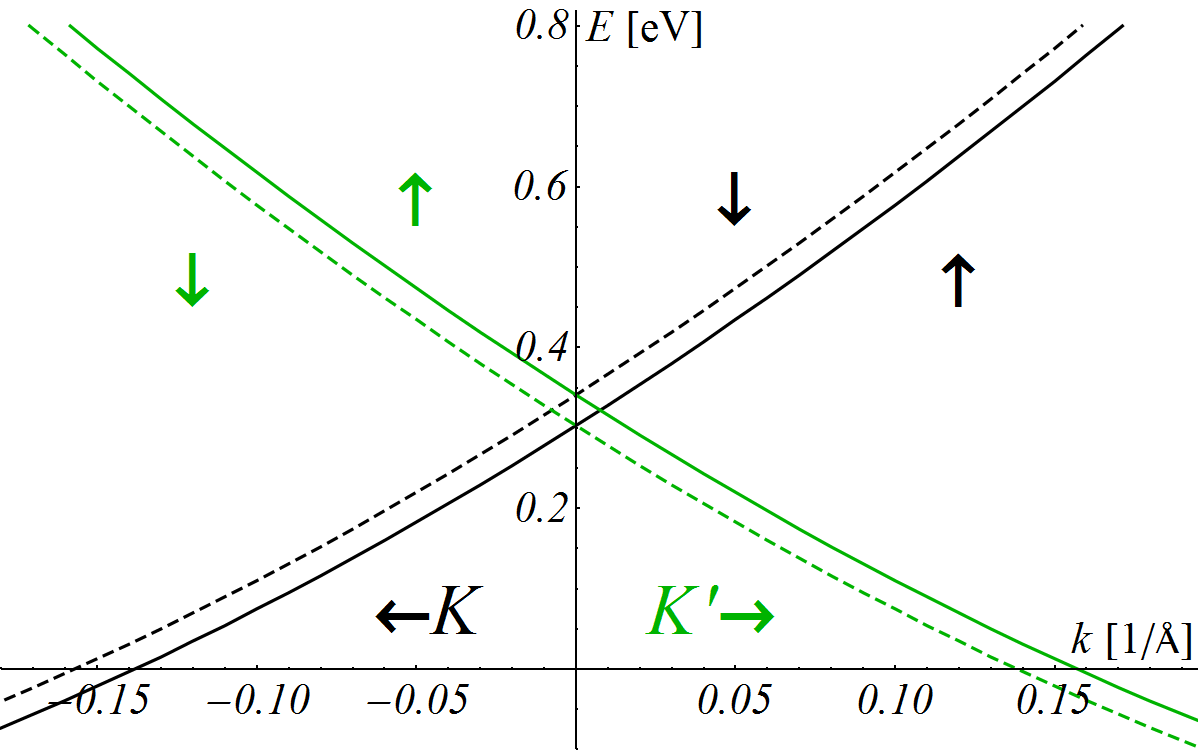}
	\caption{Band structure of the purely sulfur terminated zigzag edge in \mos\ (top edges in Fig.~\ref{fig:bs}), where electrons are propagating in opposite directions in the $K$ and $K'$ valleys. The bands are calculated for the non-magnetic state, that is the spin-splitting is only due to the bulk SOI, where the assumption $\vqm = \vqp = 1.795$ is used. Down-arrows and dashed lines denote down-spin states, up-arrows and solid lines up-spin states. The left-propagating states in the $K$ valley are in black, and the right-propagating states in the $K'$ valley are in green (gray).}
  \label{fig:spin-splitting}
\end{figure}

There is one more detail to zigzag \mos\ nanoribbons which may be potentially important. Namely, experiments have shown evidence of weak ferromagnetism in \mos\ nanostructures.\cite{zhang_magnetic_2007,tongay_magnetic_2012,gao_ferromagnetism_2013} It was suggested that this can be explained by recent calculations on zigzag \mos\ monolayer nanoribbons, which indicate that the ground state of these nanoribbons can be ferromagnetic, depending on the tensile strain,\cite{lu_strain-dependent_2012,pan_tuning_2012} and also on the edge passivation (eg.~H-saturation).\cite{vojvodic_magnetic_2009,ataca_mechanical_2011,pan_edge-dependent_2012} The consequent energy splitting of the edge states, the \textit{magnetization splitting,} depends on the average magnetic field felt by the electron in a given state. This is mainly determined by relation of the decay length of the state to the magnetic profile of the edge. Around the $K$ points, it is difficult to separate the effect of the SOI and that of the edge magnetization (EM). However, by studying the edge-bands in the relevant literature on DFT band structures,\cite{li_mos2_2008,ataca_mechanical_2011,pan_edge-dependent_2012} we can see that the combined effect of the two is more or less a rigid shift along the energy axis for the two edge-bands that can be described by our model. This shift depends on the band, but there seems to be negligible dependence on $k$ within one band, and so the change in the associated group velocities at the $K$ points are also negligibly small. Therefore, we assume that around the $K$ points, where the effective Hamiltonian $\h$ is accurate, the effect of EM on the one-dimensional edge-states can also be accounted for by a simple energy shift of the bands. To our best knowledge, the band structure of the non-magnetic state has not been published yet for such edges, although calculations were made eg.~for the total energy difference of the magnetic and non-magnetic states (see Fig.~6~of Ref.~[\onlinecite{pan_edge-dependent_2012}]). On average, this reaches only tens of meV, but some bands are affected by the EM more than others. This input would be needed to perform a proper fitting for $\vqpm_{1,2}$, after which, a simple energy-shift of the bands could restore the DFT band structures in a good approximation.
Fig.~\ref{fig:spin-splitting}~shows the bands at the purely sulfur terminated zigzag edge in the non-magnetic state, with the assumption that $\vqm = \vqp = 1.795$. The splitting at $k=0$ is 38~meV. The order of magnitude of this energy gap can be estimated from the spin-splitting of the states $\psi_\mathrm{v}$ and $\psi_\mathrm{c}$, and the ratio of their contribution to the given edge state (left part of Fig.~\ref{fig:zz-wfcn}). This splitting is significantly smaller than the values over 300~meV reported in the literature.\cite{li_mos2_2008,pan_edge-dependent_2012} This could suggest that the EM is more important than the SOI for these states, provided that the above equality of the $\vq$ angles holds. We find however that the gap changes by 1~meV for a change of 0.001 in the value of $\l$ around $\q \approx 1.795$, and for example, $\l = 0.2$ would already result in a splitting of 343~meV due only to SOI in good agreement with the DFT calculations. Nevertheless, to know exactly the relative strength of these two effects, together with the precise values of $\vqpm$, a fit to a non-magnetic band structure would be needed.
Note that this large spin-splitting combined with the fact that $K$- and $K'$-electrons are propagating in opposite directions could be used to generate spin- and valley-polarized edge currents at room temperature and above.\cite{xiao_coupled_2012} The EM can further enhance the spin-splitting for one of the valleys. In a first approximation, the magnetization is expected to shift the $K\!\!\uparrow$ band up (or down) by the same amount as the $K'\!\!\downarrow$ band shifts down (or up). A similar statement holds for the other two bands as well, but the magnetization splitting of these will be slightly different due to the valley asymmetry.

In Ref.~[\onlinecite{gibertini_emergence_2015}], the edge states of \mos\ nanoribbons are studied in relation to the width-dependent electrostatic polarization, with the help of first-principles DFT calculations. According to this study, the wider the ribbon is, the more these polarization charges are screened, but in the case of incomplete screening in narrower ribbons, a transverse electric field is induced by the polarization charges. As a result, a potential energy difference appears between the edges, and the edge bands are pushed upward at one side of the ribbon, and pushed downward at the other side. This phenomenon is independent of the spin state, but otherwise, in our model, it can be accounted for in a similar fashion as EM, with a rigid shift of the bands.

Recently, several works\cite{klinovaja_spintronics_2013,xiao_coupled_2012} used a ``gapped-graphene'' type $\scp kp$ Hamiltonian to model the bulk electronic structure of monolayer \mos\ in the vicinity of the $K$ ($K'$) points. While such analogy to graphene can be useful to understand bulk properties, the analogy breaks down at the edges. To highlight the difference to gapped graphene, it is instructive to consider a zigzag edge in the limit of vanishing or negligible $\A$ and $\B$ parameters and equal energy gaps for the two valleys and spins. Here we assume that the restored electron--hole symmetry of the bulk is not broken by the edge. From this, it follows that $\q=\vq=\pi/2$, and we obtain edge states with Dirac-like, linear dispersion $\ve=\pm \G \kl$ for bands belonging to the $K$ and $K'$ valleys respectively. The normal component $\kp=i\Eg/(2\G)$ depends only on material parameters. These results are independent from the edge geometry, that is all zigzag edges are equivalent in the model as long as $\A=\B=0$. Differences between the edges can be taken into account only if finite values for $\A$ and $\B$ are allowed. But why can we not find these in-gap states in gapped graphene?\cite{yao_edge_2009} To answer this, we recall that the basis of graphene is a sublattice basis, in which the edge inherently breaks the electron--hole symmetry, since all the terminating sites belong to one particular sublattice, and all have the same on-site energy, which is also the energy of the flat bands (see Fig.~1~of Ref.~[\onlinecite{yao_edge_2009}]). In \mos, the basis functions are extended wave functions of the valence- and conduction-bands, and the terminating sites do not belong to either of these. Therefore, it is conceivable that none of the two basis functions is given preference by the edge, thus not causing any symmetry breaking. Based on these findings, we expect the presence of edge states showing the preserved electron--hole symmetry of the gapped system.

In the case of an armchair ribbon, provided there is no asymmetric edge reconstruction, we have the same kind of edges on both sides: $\l_1=\l_2$ and $\q_1=\q_2$. We have two different linear combinations of states localized to the opposite edges, and these combinations are split in energy, but they are degenerate in spin. In narrow ribbons, where the edge states do not vanish completely at the opposite edges, the difference $\Delta\f=\f_2-\f_1$ affects the band structure, but the angles $\f_{1,2}$ individually do not. In wide ribbons, where the opposite edges do not interfere, the angles $\f$ have no effect on the dispersion, the splitting vanishes, and we end up with fourfold degenerate edge states: twice for the spin and twice for the two linear combinations, where the latter ones can eventually be split into states belonging to their respective edges only. Here, the angles $\f$ turn out to be arbitrary phase differences between the valleys, as in $\{\psi_\mathrm{v},\psi_\mathrm{c},-\psi_\mathrm{c}'\exp(i\f),\psi_\mathrm{v}'\exp(i\f)\}$, and this phase difference does not change upon time reversal.

It is apparent from Fig.~3(b) of Ref.~[\onlinecite{erdogan_transport_2012}] that the band structure of the edge states does not change qualitatively with the width in the given range, and the valley-mixing effect of $\Delta\f$ can be neglected. Because of the size of our basis, we cannot describe all the bands, but only one of the degenerate flat bands around the $K^*$ point (Fig.~\ref{fig:kspace}), denoted by $\Gamma$ in Fig.~3(b) of Ref.~[\onlinecite{erdogan_transport_2012}]. For small values of $\l$, almost in the full range of $\q$, we see a flat band indeed that could be matched with any of the two between 0 and 1~eV. Without further input however, eg.~regarding the composition of these states, we cannot tell which one to fit to.

Note however that in graphene nanoribbons, we can also describe extended edge-to-edge bulk states with the $M$ matrix method, where $\Delta\f$ also has an effect on the dispersion of these states. This can be used to determine its value eg.~by means of a TB model, which enables the analytical determination of band structure, like it has been done in Ref.~[\onlinecite{wakabayashi_electronic_2010}]. For the spectrum at the $K$ point, these authors found $\ve_r = t \left|1+2\cos\frac{r \pi}{N_\mathrm{d}+1}\right|$, where $t$ is the transfer integral between nearest-neighbor carbon sites, and $r\in\{1\dots N_\mathrm{d}\}$ is an integer indexing the bands. For the lowest, positive energy $\ve_{r_0}$, $\left|\frac{r \pi}{N_\mathrm{d}+1}-\frac{2\pi}{3}\right|$ is minimal with $r=r_0$. It can also be shown easily that $\Delta\f=\pm 2 W \varepsilon_{r_0}/\G$ holds, where $W=\frac{1}{2}N_\mathrm{d} a$, and $\G=\frac{\sqrt{3}}{2}t a$. Combining all this, we obtain that $\Delta\f=\pm \frac{2}{3}\pi$ if $N_\mathrm{d}\ \mathrm{modulo}\ 3$ is 0 or 1, and $\Delta\f=0$ if $N_\mathrm{d}\ \mathrm{modulo}\ 3$ equals 2. The sign of $\Delta\f$ is irrelevant for armchair nanoribbons because of their symmetric edges, this is why we cannot find a definite sign from the energy spacing. Nevertheless, it is sensible to think of this effect as if every additional dimer added $2\pi/3$ to $\f_2$ counting from the narrowest ribbon with $N_\mathrm{d}=3$, $\f_1=0$ and $\f_2=2\pi/3$. As for the in-gap states in TMDC armchair nanoribbons, we speculate that $\Delta\f$ may be determined by the width of the ribbon as well. The main arguments are that the individual angles $\f_1$ and $\f_2$ do not affect the in-gap bands, but only their difference, and the effect of $\Delta\f$ vanishes with the vanishing overlap of opposite lying edge states in an infinitely wide ribbon. These findings suggest that the relative geometry of the edges matter, as in the case of graphene, where $\Delta\f$ is directly proportional to the ribbon width $W$. The determination of the exact relation between $\Delta\f$ and $W$ in TMDCs is an open problem, which could probably be solved by a TB calculation.\cite{liu_three-band_2013,cappelluti_tight-binding_2013,rostami_effective_2013,chu_spin-orbit-coupled_2014,rostami_valley_2015,fang_ab-initio_2015}

\section{Conclusions}

We derived the general boundary conditions for zigzag and armchair edges of transition-metal dichalcogenide monolayers within a continuum model. We used an effective, two-band, $\scp kp$ Hamiltonian around the K point, where the electron--hole symmetry breaking quadratic terms were taken into account. We modeled the effect of the edges with the $M$ matrix method, and after reducing the number of free parameters by symmetry considerations, we found that different edge geometries can be described with two or three scalar parameters varying between 0 and $2\pi$. We discussed the case of single edges and interacting edges of nanoribbons, and we fitted the free parameters to \textit{ab initio} band structures. Focusing mainly on zigzag edges, we analyzed the edge states and their dispersion relation in \mos\ in particular, and we demonstrated a good agreement with DFT based tight binding calculations.

Note that our method is not restricted to in-gap states, but it can be applied to low-energy bulk states as well, as long as they are close to the conduction band or the valence band edge. Therefore our method can be used to study bulk band states in edge-terminated nano-structures, such as quantum dots and nanoribbons.\cite{brey_electronic_2006} This also opens up the possibility to fit the $\vqpm_{1,2}$ angles for the two edges using these extended bulk states. The advantage of this would be that the bulk bands are not much affected by the EM in wider ribbons, but the disadvantage is that the bulk states connect the edges and a simultaneous fitting for $\vqpm_1$ and $\vqpm_2$ is needed. The ribbons cannot be too wide either so that the quantized bulk bands are well-separated in energy. Alternatively, as demonstrated in our work, fitting to non-interacting in-gap edge states is the simplest way to obtain $\vqpm_{1,2}$, but for accuracy, non-magnetic band structures need to be calculated first.

\acknowledgments

We are grateful for the support of Gotthard Seifert, who provided us with the nanoribbon coordinates and the DFT data from Ref.~[\onlinecite{erdogan_transport_2012}], and we also acknowledge fruitful discussions with Marco Gibertini. This work was supported by the DFG program SFB767.

%

\end{document}